\documentclass[a4paper]{spie}  
\usepackage[]{graphicx}

\title{Gaia: focus, straylight and basic angle} 

\author{
A.~Mora\supit{a}\supit{b},
M.~Biermann\supit{c},
A.~Bombrun\supit{a}\supit{d},
J.~Boyadian\supit{e},
F.~Chassat\supit{e},
P.~Corberand\supit{e},
M.~Davidson\supit{f},
D.~Doyle\supit{g},
D.~Escolar\supit{g},
W.L.M.~Gielesen\supit{h},
T.~Guilpain\supit{e}\supit{i},
J.~Hernandez\supit{a},
V.~Kirschner\supit{g},
S.A.~Klioner\supit{j},
C.~Koeck\supit{e},
B.~Laine\supit{g},
L.~Lindegren\supit{k},
E.~Serpell\supit{l}\supit{m},
P.~Tatry\supit{e},
and P.~Thoral\supit{e}
\skiplinehalf
\supit{a}ESA-ESAC Gaia Science Operations Centre, Camino Bajo del Castillo s/n, Urb. Villafranca del Castillo, 28692 Villanueva de la Ca\~{n}ada, Madrid, Spain; \\
\supit{b}Aurora Technology, Crown Business Centre, Heereweg 345, 2161 CA Lisse, The Netherlands; \\
\supit{c}Astronomisches Rechen-Institut, Moenchhofstr. 12-14, 69120 Heidelberg, Germany; \\
\supit{d}HE Space Operations GmbH, Flughafenallee 24, 28199 Bremen, Germany; \\
\supit{e}Airbus Defence and Space, 31 rue des Cosmonautes, Z.I. du Palays, 31402 Toulouse Cedex 4, France; \\
\supit{f}Royal Observatory, Blackford Hill View, Edinburgh EH9 3HJ, United Kingdom; \\
\supit{g}ESA-ESTEC Directorate of Technical and Quality Management, Keplerlaan 1, 2201 AZ Noordwijk, The Netherlands; \\
\supit{h}TNO Science and Industry, Stieltjesweg 1, 2600 AD Delft, The Netherlands; \\
\supit{i}Altran Technologies, 17 Avenue Didier Daurat, 31700 Blagnac, France; \\
\supit{j}Lohrmann Observatory, Dresden Technical University, Mommsenstr. 13, 01062 Dresden, Germany; \\
\supit{k}Lund Observatory, Department of Astronomy and Theoretical Physics, Lund University, Box 43, 22100, Lund, Sweden; \\
\supit{l}ESA-ESOC Gaia Flight Control Team, Robert-Bosch-Strasse 5, 64293 Darmstadt, Germany; \\
\supit{m}Telespazio VEGA Deutschland GmbH, Europaplatz 5, 64293 Darmstadt, Germany; \\
}

\authorinfo{Further author information: (Send correspondence to A.M.)\\A.M.: E-mail: alcione.mora@esa.int, Telephone: +34 91 813 1480}

\begin{document} 
  \maketitle 

\begin{abstract}
The Gaia all-sky astrometric survey is challenged by several issues affecting the spacecraft stability. Amongst them, we find the focus evolution, straylight and basic angle variations

Contrary to pre-launch expectations, the image quality is continuously evolving, during commissioning and the nominal mission. Payload decontaminations and wavefront sensor assisted refocuses have been carried out to recover optimum performance.
Straylight and basic angle variations several orders of magnitude greater than foreseen were found and studied during commissioning by the Gaia scientists (payload experts). Building on their investigations, an ESA-Airbus DS working group was established during the early nominal mission and worked on a detailed root cause analysis. In parallel, Gaia scientists have also continued analysing the data, most notably comparing the BAM signal to global astrometric solutions, with remarkable agreement.

In this contribution, a status review of these issues will be provided, with emphasis on the mitigation schemes and the lessons learned for future space missions where extreme stability is a key requirement.
\end{abstract}




\section{Introduction}

The ESA Gaia mission is creating the most comprehensive 3D census of the Galaxy ever envisaged, comprising more than a billion parallaxes and proper motions, complemented with exquisite visible spectrophotometry and millions of radial velocities, providing astrophysical parameters, non-single star solutions, solar system objects, variability, etc. \cite{2014EAS....67...15P}. See also Prusti et al. (2016 A\&A, in press).

The most important quality behind the Gaia overall design is its extreme stability. In this work, some issues related to that stability are reviewed, both to explain the Gaia performance and as a source of information for future missions.

The general architecture and design choices needed to obtain very high stability from purely passive means are described in Sect.~\ref{sect:architecture}. The evolution of the telescopes focus is presented in Sect.~\ref{sect:focus}. The analysis of the additional straylight is given in Sect.~\ref{sect:straylight}. The basic angle variations are discussed in detail in Sect.~\ref{sect:basicAngle}. Finally Sect.~\ref{sect:conclusions} provides the conclusions.

\section{Gaia architecture and stability}
\label{sect:architecture}

The main force driving the Gaia design and architecture is to achieve maximum stability by purely passive means. In this way, there is no active thermal or mechanical control neither within the payload nor the service module, except for survival purposes.

\begin{figure}
\begin{center}
\includegraphics[width=0.49\hsize]{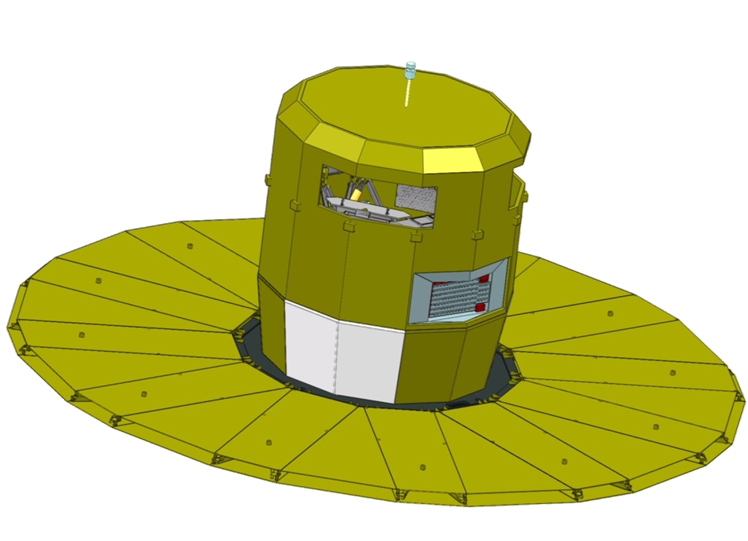}
\includegraphics[width=0.49\hsize]{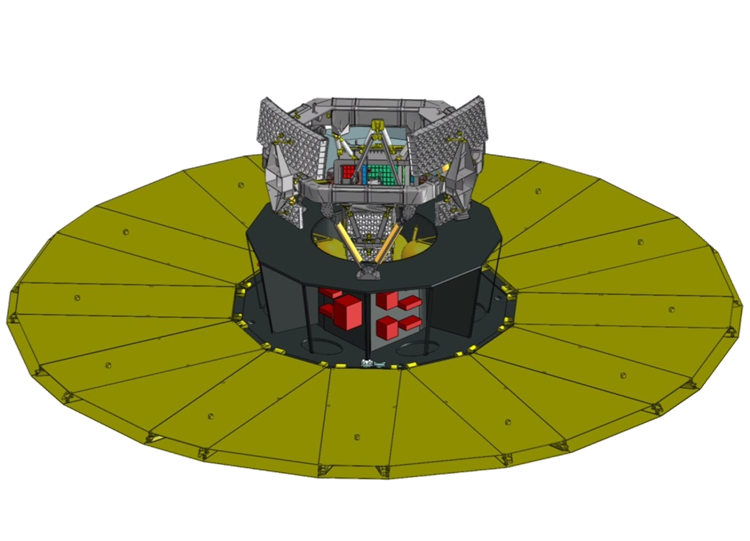}
\end{center}
\caption{Gaia architecture layout. Left: The thermal tent, service module and deployable sun-shield are apparent. The Sun would be downwards, keeping 45$^\circ$ aspect angle with respect to the spin axis. Right, the payload is located within the thermal tent and atop the service module, isolated from it via three double bipods. Images: Airbus DS.
\label{fig:architecture}}
\end{figure}

The spacecraft architecture is structured to provide maximum isolation of the payload module with respect to the ambient changes, see Fig.~\ref{fig:architecture}. The payload is composed of two telescopes, a focal plane and a spectrometer mounted on a circular optical bench, the torus. The latter is connected to, but isolated from the service module via three bipods, which have double struts: carbon fibre and glass fibre reinforced polymers. The carbon fibre struts provide rigidity and resistance against the launch forces, but have high thermal conductivity. They are thus decoupled after launch, leaving the good insulator glass fibre struts as the main interface between the service and payload modules. 

The service module is composed of a main body, a thermal tent and a deployable sun shield. The main body provides structural integrity through a CFRP truncated cone, support for the equipment (communications, propulsion, electric power, computers, atomic clock, ...) and serves as base for the payload. The thermal tent encapsulates the payload module and provides further stability. The deployable sun shield avoids direct illumination of the payload while the spacecraft rotates around the spin axis at a constant sun aspect angle of 45$^\circ$, which results in a constant irradiation. It also provides an appropriate operational temperature of $\sim -110^\circ$C for the detectors.

The general principle is that Gaia has no moving parts, which means using laser gyros instead of mechanical units, and a phase array beam steering antenna, instead of mechanical joints and a moving dish. The only exception is the focusing mechanism, see Sect.~\ref{sect:focus}, which is very rarely operated. All components susceptible to consume significant and variable amounts of power, such as the computers, transponders and even the atomic clocks, are located in the service module.

\section{Focus evolution}
\label{sect:focus}

\begin{figure}
\begin{center}
\includegraphics[width=0.75\hsize]{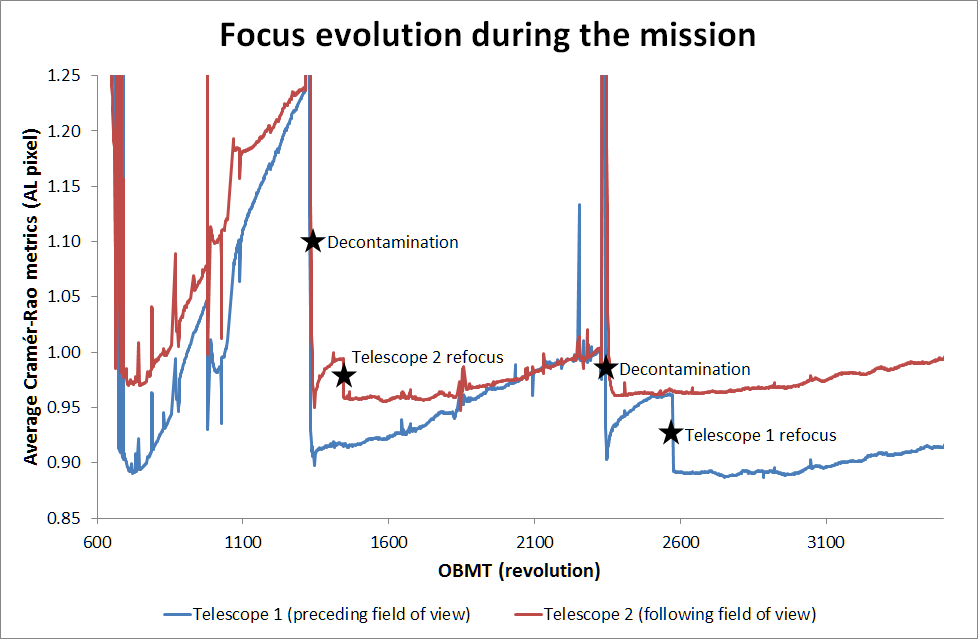}
\end{center}
\caption{Focus evolution during the mission. The average Cram\'er-Rao image sharpness metrics is plotted for each telescope from early commissioning (March 2014) to March 2016. Major payload decontaminations and telescope refocuses are indicated. 
\label{fig:cramerRao}}
\end{figure}

Gaia needs an almost diffraction-limited optical quality throughout the field of view of each telescope to provide exquisite astrometry. The key metrics is the Cram\'er-Rao image sharpness \cite{1978moas.coll..197L} \cite{2014SPIE.9143E..0XM}, which for a given LSF is expressed as:

\begin{equation}
  \sigma_\eta = \frac{1}{\sqrt{ \displaystyle\sum_{k=0}^{n-1} \frac{ (S'_k)^2 }{r^2 + b + S_k} }}
\end{equation}

where $S_k$, the LSF, is the number of electrons collected from the star, binned Across Scan (AC), for Along Scan (AL) pixel coordinate $k$, where $k \in [0, n-1]$. $S'_k$ is the derivative of $S_k$ with respect to the pixel coordinate, $r$ the read-out noise (in electrons) and $b$ the homogeneous sky background (in electrons). The units of $\sigma_\eta$ are pixels. $\sigma_\eta$ is a measure of the astrometric information present in a given LSF: it is the maximum centroiding precision that an optimum maximum-likelihood method can achieve. If only bright stars are considered, $b$ is subtracted and $\sigma_\eta$ is normalised by the photon noise:

\begin{equation}
  {\rm Cramer-Rao}_{\rm normalised}
   = \sigma_\eta \sqrt {\displaystyle\sum_{k=0}^{n-1} S_k}
   = \sigma_\eta \sqrt {N_{e^-}}
\end{equation}

which is independent of the stellar brightness, and can be averaged over the whole focal plane. Fig.~\ref{fig:cramerRao} shows the evolution of the Cram\'er-Rao image sharpness throughout the mission. Several things are apparent. First, the image quality is very good, being always around 1.0, as expected for the nearly diffraction-limited and slightly undersampled Gaia PSF. Second, telescope 1 provides better AL image quality\cite{2014SPIE.9143E..0XM}. Third, decontamination is always beneficial in terms of image quality, although small focus adjustments are typically needed afterwards to regain optimum performance. Fourth, the focus has never been stable, although the degradation slope becomes shallower as the mission evolves. The root cause has not been identified, although a number of hypothesis have been postulated: glue shrinkage, hysteresis, water contamination, etc.

\begin{figure}
\begin{center}
\includegraphics[height=5cm]{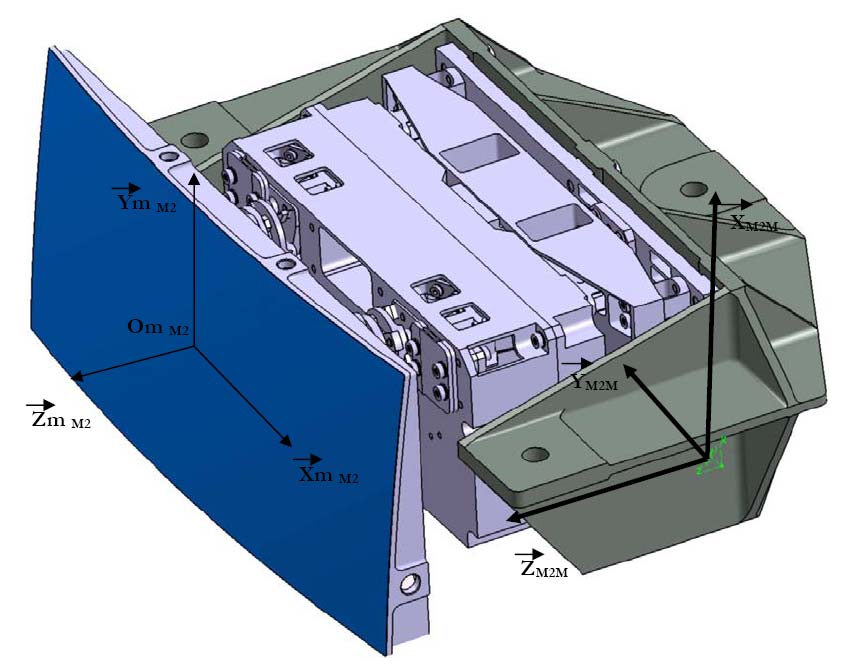}
\includegraphics[height=5cm]{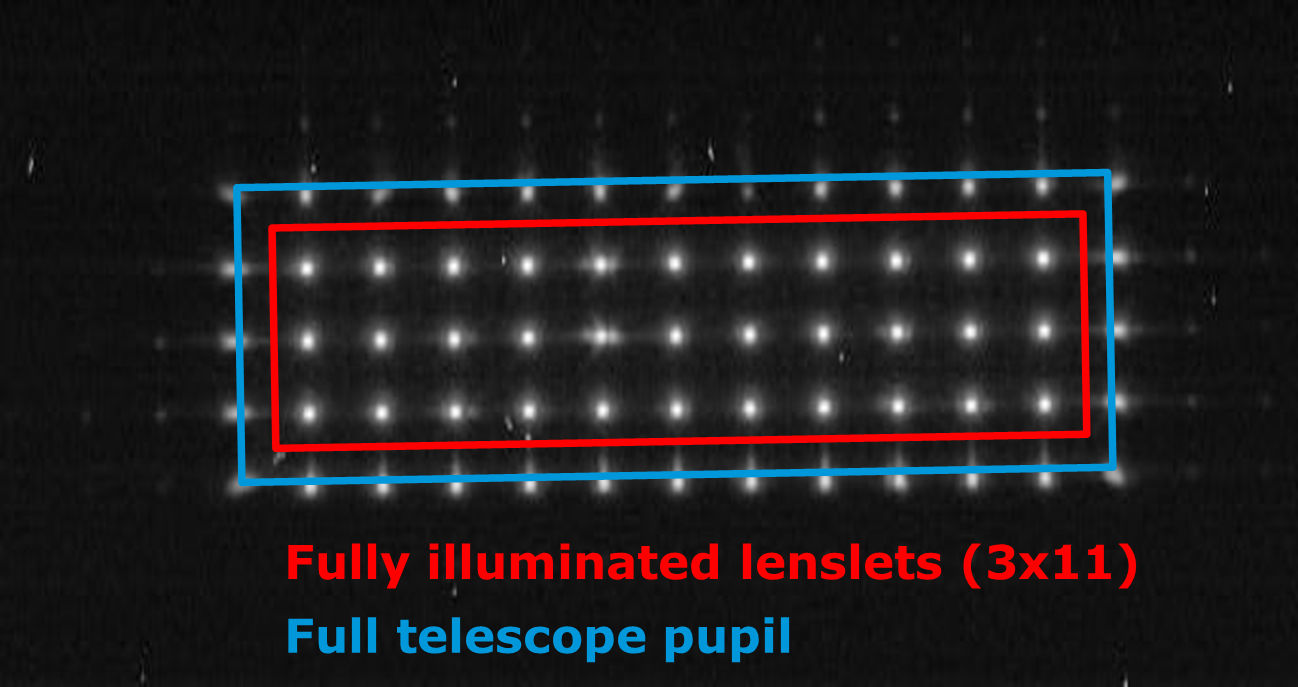}
\end{center}
\caption{M2MM overview (left) and WFS sample pattern (right). The M2MM provides five degrees of freedom (three translations and two rotations) control over the Gaia telescope secondary mirrors. They are only operated during the rare refocusing events. Image: Airbus DS. Two Shack-Hartmann WaveFront Sensors (WFS) provide low-order sampling needed during the first focusing attempts in early commissioning and differential measurements for fine tuning during the nominal mission. 
\label{fig:m2mmWfs}}
\end{figure}

These rare but important focus adjustments are the only times the M2 Movement Mechanisms (M2MM) are operated\cite{2009SPIE.7439E..14V}. Basically, they provide five degrees of freedom (three translations and two rotations) control over the secondary mirrors, which has been sufficient to obtain a superb image quality. Two WaveFront Sensors (WFS) provide key information to avoid blind exploration of the 5-D actuation parameter space. The pre-launch preparations and its use during early commissioning are described in\cite{2012SPIE.8442E..1QM}\cite{2014SPIE.9143E..0XM}. Basically, they were used in absolute mode, the zero wavefront reference being provided by a trio of reference fibres. These readings were essential to recover a reasonably sharp PSF after launch. See Fig.~\ref{fig:m2mmWfs}, adapted from\cite{2014SPIE.9143E..0XM}, for an overview of the M2MM and a WFS pattern.

However, the WFS have a small number of microlenses (pupil sampling vs stellar brightness trade-off), which induces a small aliasing during the Legendre polynomial decomposition. In addition, the minimum wavefront point is just a good (and close) starting point for the optimum balanced configuration between all focal planes (astrometric, photometric and spectroscopic). The so-called ``best focus'' was a trade-off configuration iteratively obtained after all scientific data were analysed\cite{2014SPIE.9143E..0XM}.

Once an optimum configuration was defined, the WFS do not need to operate in absolute mode any more, but just determine the differences between the current situation and best focus. In addition, the most sensitive actuation is, of course, pure M2 $z$-axis focus. Therefore, the two focus corrections that have taken place during the nominal mission have been small pure $z$-axis focus adjustments (+2~$\mu$m in both cases). Optimum performance has been regained afterwards. This strategy is thus the baseline for the remaining of the Gaia mission: payload decontamination followed by small $z$-axis refinements, when needed.

\section{Straylight}
\label{sect:straylight}

\begin{figure}
\begin{center}
\includegraphics[width=0.49\hsize]{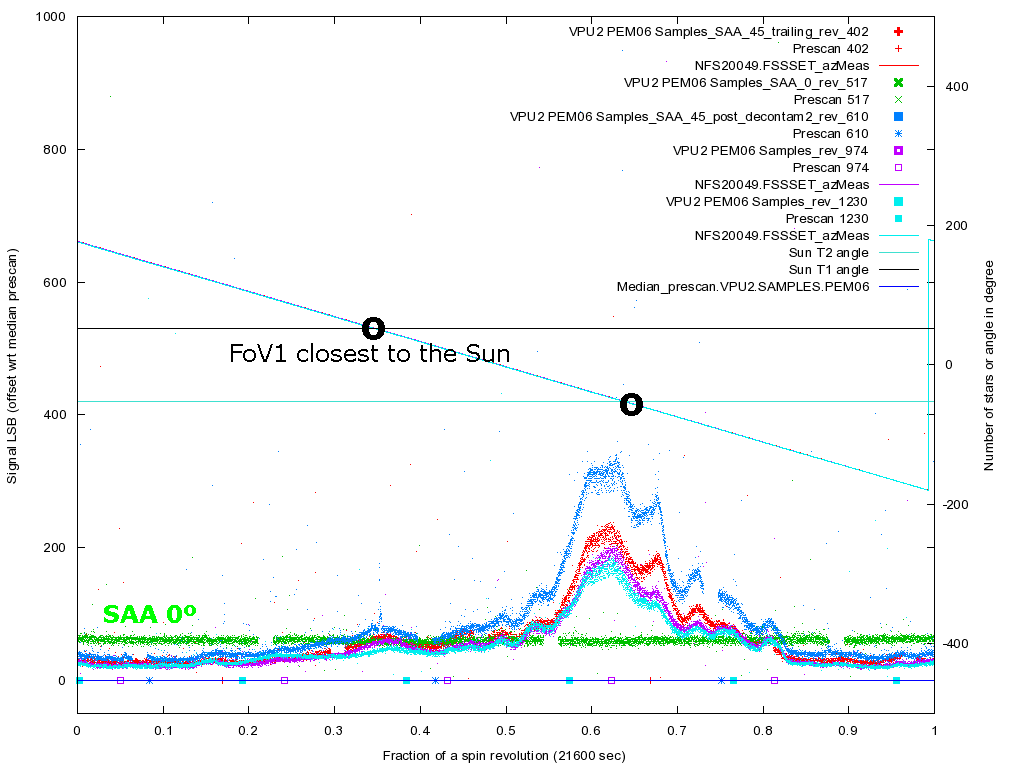}
\includegraphics[width=0.49\hsize]{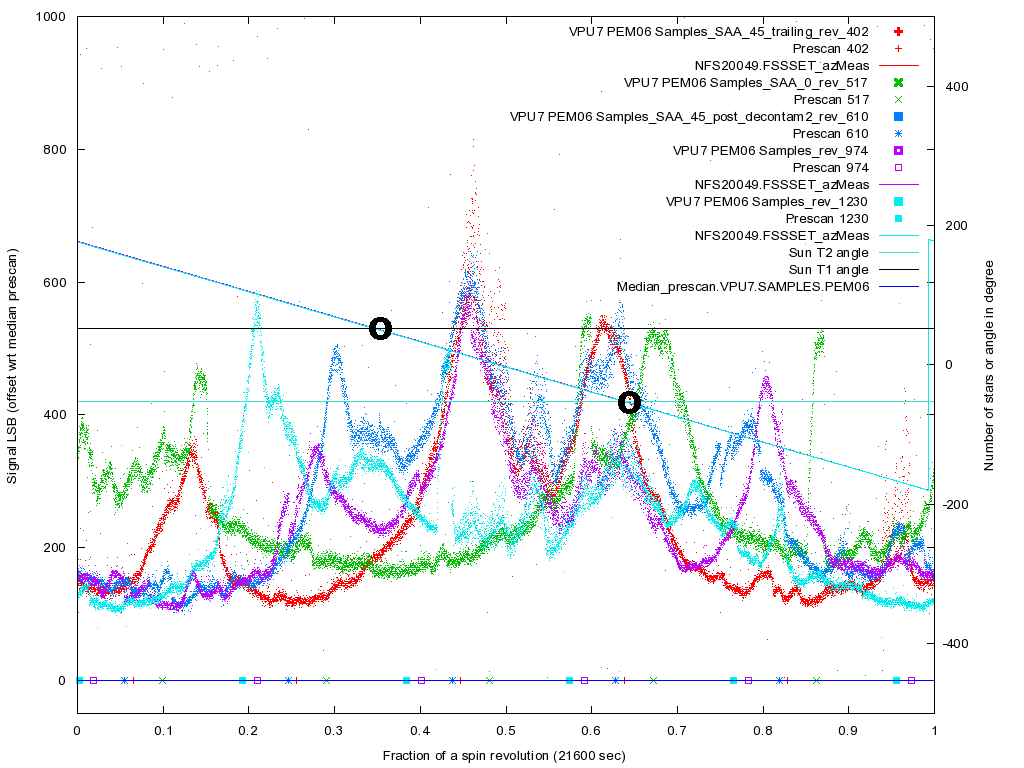}
\end{center}
\caption{Straylight evolution for selected spacecraft revolutions. Left: VPU2, PEM06. Right: VPU7, PEM06. The $x$-axis shows the heliotropic spacecraft spin phase. The times when the Sun is closer to telescopes 1 and 2 are indicated by black circles. The left panel shows a CCD whose main straylight contributor is the Sun, while additional sources, mostly the Galaxy, are also a concern for other detectors. Images: Airbus DS.
\label{fig:straylightHeliotropicPhase}}
\end{figure}

A significantly greater than expected straylight was discovered during early commissioning by the Gaia scientists (payload experts). Lots of effort from their side, together with ESA and Airbus DS, were invested in both understanding its origin and minimising its effects. A number of tests were carried out, most notably operating Gaia at non-nominal Sun aspect angles of 42$^\circ$ and 0$^\circ$, while collecting stellar data. Fig.~\ref{fig:straylightHeliotropicPhase} shows the straylight levels for two CCDs as a function of the heliotropic rotation phase during selected spacecraft rotations.

Two distinct behaviours are found. For some CCDs, such as VPU2, PEM06, the main straylight peak always takes place when the telescope 2 aperture is closer to the Sun, which converts to a flat non-zero level for a Sun aspect angle of 0$^\circ$. This is consistent with a pure solar origin. However, other detectors show additional peaks in addition to that of purely solar origin. This is evident by the behaviour at Sun aspect angle 0$^\circ$.
 
\begin{figure}
\begin{center}
\includegraphics[height=4.3cm]{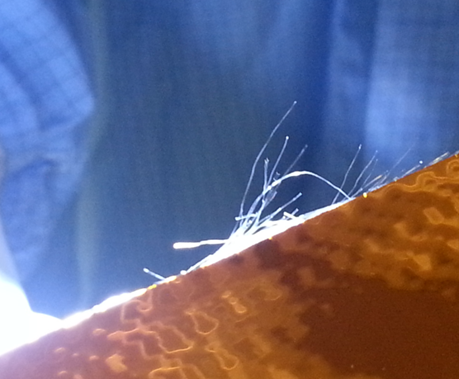}
\includegraphics[height=4.3cm]{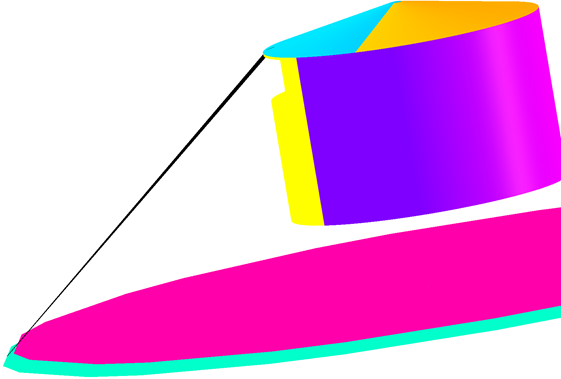}
\includegraphics[height=4.3cm]{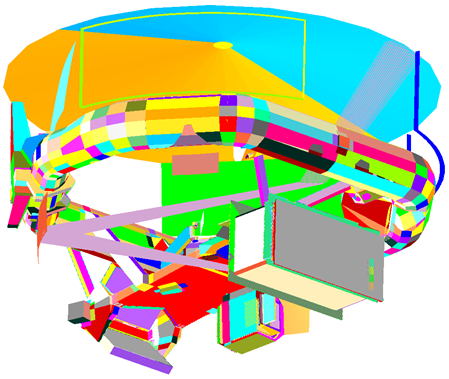}
\end{center}
\caption{Origin of the solar straylight. It is the combination of three effects: Nomex fibres sticking out of the sun-shield blankets (left), the sun-shield design allowing some diffracted light to get through the thermal tent telescope apertures (centre) and rogue paths transporting that scattered light into the focal plane (right). 
\label{fig:straylightSun}}
\end{figure}

The solar contribution was completely unexpected. Different tests were carried out to determine its origin. Most notably, a gentle dependence with the Sun aspect angle was found, at odds with simple diffraction models of a two layer sun-shield. 
Understanding of this phenomenon was one of the mandates of the ESA-Airbus DS basic angle variations and straylight working group, which carried out the investigations after commissioning. The ESA-ESTEC team within the group finally found the origin, which is the (unfortunate) combination of three effects. First, there are sticking out Nomex fibres at some edges of the sun shield (the untapered triangular sections). the fibres are needed for mechanical integrity. The ends protruding from the blanket edges were a known fact before launch, but not considered a major risk from the mechanical and thermal points of view (the main drivers behind the sun-shield development). Second, the sun-shield design allows some diffracted light from the first blanket to illuminate the upper part of the thermal tent apertures (no direct illumination, though). Third, there are some unbaffled rogue paths that can transport that scattered light from the top part of the thermal tent down to the focal plane. See Fig.~\ref{fig:straylightSun} for an overview of the three effects.

\begin{figure}
\begin{center}
\includegraphics[width=0.85\hsize]{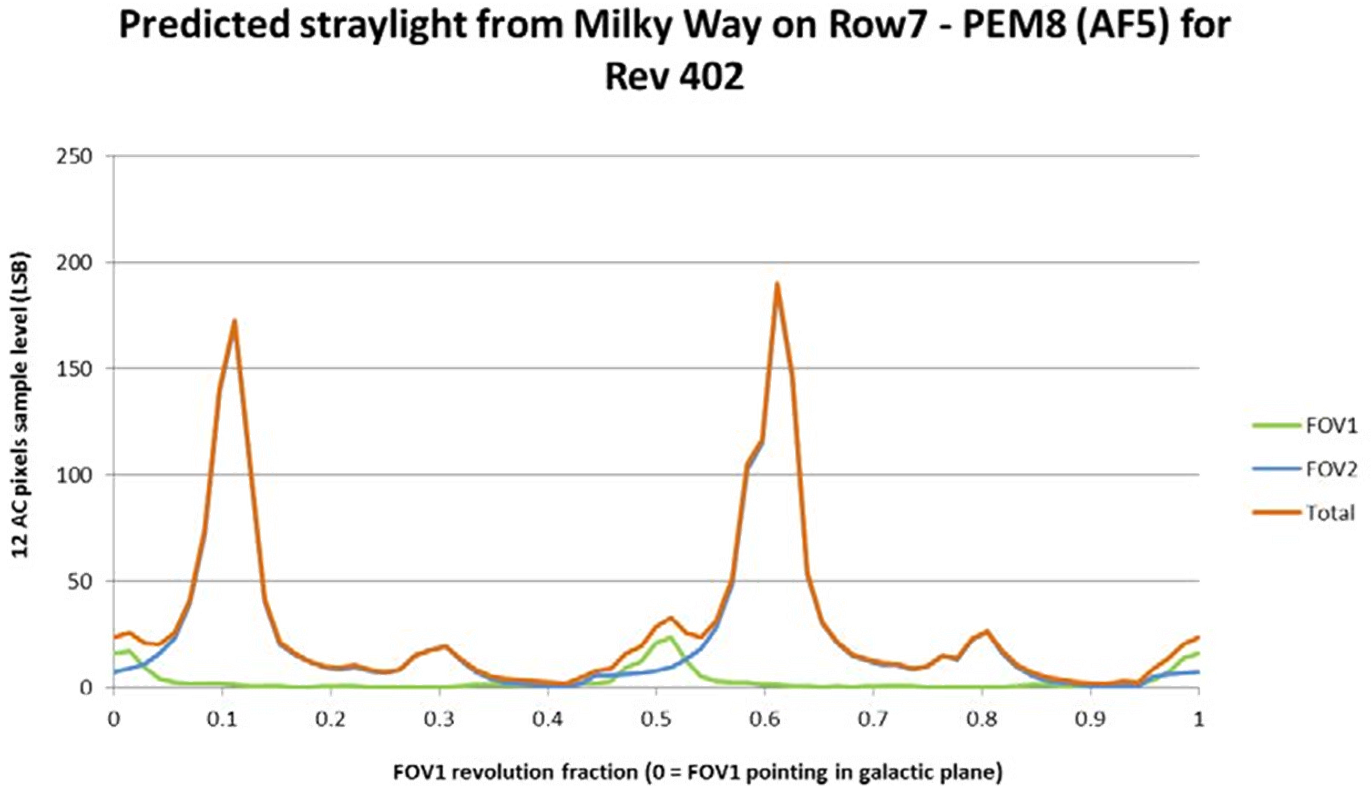}
\end{center}
\caption{Straylight generated by the Galaxy. The Milky Way brightness distribution, when convolved with the straylight paths within the payload, can produce a significant straylight signature, in agreement with the in-orbit measurements. Plot: Airbus DS.
\label{fig:straylightGalaxy}}
\end{figure}

Regarding the non-solar straylight components. A detailed straylight analysis was carried out before launch. Some paths were identified, but they were found harmless against typical bright stars. However, it was realised by Airbus DS during commissioning that the integrated light of the whole Galaxy, when convolved with the straylight paths, could produce a measurable signature, in agreement with the in-orbit measurements. See Fig.~\ref{fig:straylightGalaxy} for an example calculation.

Once the origin of the straylight was identified, it was realised that it could not be avoided. Mitigation measures have thus been undertaken. Mostly, the data collection magnitude threshold is now adaptive for the radial velocity spectrometer, which is now straylight Poisson noise driven, as opposed to the pre-launch CCD read-out noise limited expectations. This means the faint object detection limit is now shallower during the maximum solar straylight peaks. In this way, only useful data are downlinked. Additional on-board functionalities have also been implemented, but not yet used, such as adaptive AC bin size for the radial velocity spectrometer or a smaller AC bin size for the faintest stars in the astrometric field. Finally, the on-ground downstream processing has also been improved to face the additional noise in all focal planes. The end-of-mission performance has been updated accordingly.

\section{Basic angle variations}
\label{sect:basicAngle}

The analysis of the basic angle variations comprises the bulk of this work. Sect.~\ref{sect:basicAngleOverview} provides an overview of the basic angle, its meaning importance and measurement. Sect.~\ref{sect:bamFringePhasePeriod} describes the current knowledge of the BAM fringe phase and period variations and discontinuities, Selected work of the ESA-Airbus DS basic angle and straylight working group are presented in Sect.~\ref{sect:basicAngleWorkingGroup}. Finally, a detailed fringe by fringe analysis and its application to the search for the white fringe are shown in Sect.~\ref{sect:bamShapeWhiteFringe}.

\subsection{Basic angle overview}
\label{sect:basicAngleOverview}

The basic principle of Gaia has been described several times\cite{2011EAS....45..109L}. See also Prusti et al. (2016 A\&A, in press). To obtain absolute parallaxes, the differential positions in stars observed in fields of view separated by a large angle are determined by simultaneous observations with two telescopes whose line of sight is separated by a fixed and large basic angle. When a significantly large number of (pairs of) measurements is acquired, with different orientations on the sky, the individual proper motions and parallaxes can be determined for each star in an iterative way.

However, deriving absolute parallaxes relies in the basic angle either being stable, or at least known, at levels below the Gaia accuracy floor, which is a fraction of $\mu$as. Low frequency variations, slower than the six hours rotation period, are easily accounted for by self-calibration. Very short period variations are averaged during all transits. Heliotropic angle rotation-synchronised systematic variations provide the greatest challenge. They can create systematic errors in the astrometry (e.g. a shift in the parallax zero point) and it is unclear whether they can be fully self-calibrated (there are ongoing efforts in this direction, though).

\begin{figure}
\begin{center}
\includegraphics[height=5.4cm]{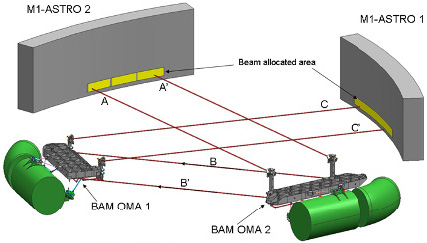}
\includegraphics[height=5.4cm]{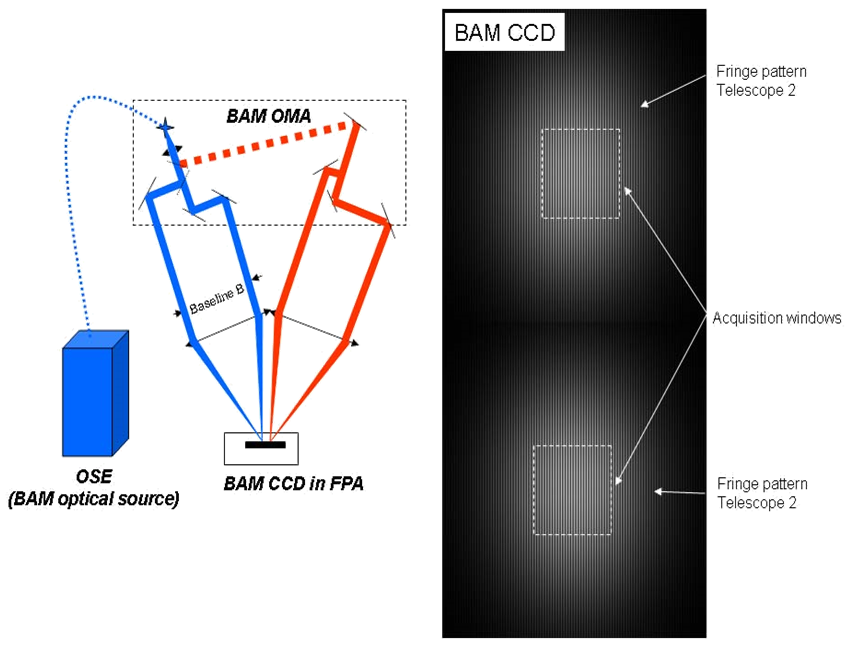}
\end{center}
\caption{Basic Angle Monitor overview. Left and Middle: the BAM is composed of two optical benches that divide the light from a single laser source into four beams, injecting two of them into each telescope entrance pupil. Right: for each telescope, the two input laser beams create a Young interference pattern in the focal plane. They are just very high signal to noise artificial stars, whose relative movement traces changes in the basic angle. Images: Airbus DS.
\label{fig:bamOverview}}
\end{figure}

An on-board metrology system, the Basic Angle Monitor (BAM), was included in the payload as a means to ensure the mission goals can be achieved even in the presence of such Sun synchronous perturbations. The BAM works by differential measurements. An artificial star is projected through each telescope onto a common dedicated detector in the focal plane. If the relative AL distance between the artificial stars change, this means the basic angle has varied. In order to achieve high precision in a short time, the artificial stars are interference patterns generated from a single laser source divided into four beams, two per telescope. An overview of the whole system is presented in Fig.~\ref{fig:bamOverview}

The design rules guaranteeing the BAM measurements are related to real basic angle variations and not to instabilities in the BAM itself are described by \cite{2014SPIE.9143E..0XM}. It also presents the precision requirements, which are a differential measurement in the along scan direction better than 0.5~$\mu$as each 10 minutes. The on-board behaviour indeed fulfils those requirements, which are absolutely unprecedented. Note that if such a tiny angular shift is interpreted in terms of primary mirror rotations, we are considering pm level displacements, equivalent to microfringe shifts or, in terms of the SiC crystal structure, subatomic movements ($\sim$0.02 atoms). In terms of noise requirements, the system provides 12.6~$\mu$as~Hz$^{-1/2}$	$\sim$ 34~pm~Hz$^{-1/2}$ in a frequency range starting at 0.043~Hz, and well below the sub-mHz regime, ideally up to 2.3e-5~Hz. This is even better than the eLISA gravitational waves mission mHz requirements \cite{2013GWN.....6....4A}.

\subsection{BAM fringe phase and period variations and discontinuities}
\label{sect:bamFringePhasePeriod}

\begin{figure}
\begin{center}
\includegraphics[width=0.49\hsize]{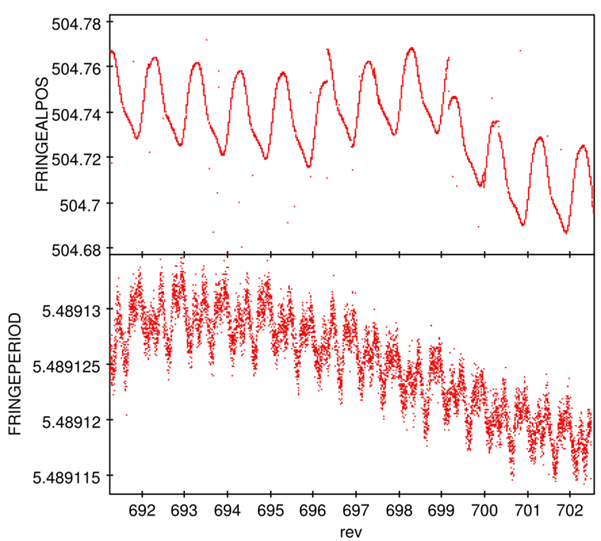}
\includegraphics[width=0.49\hsize]{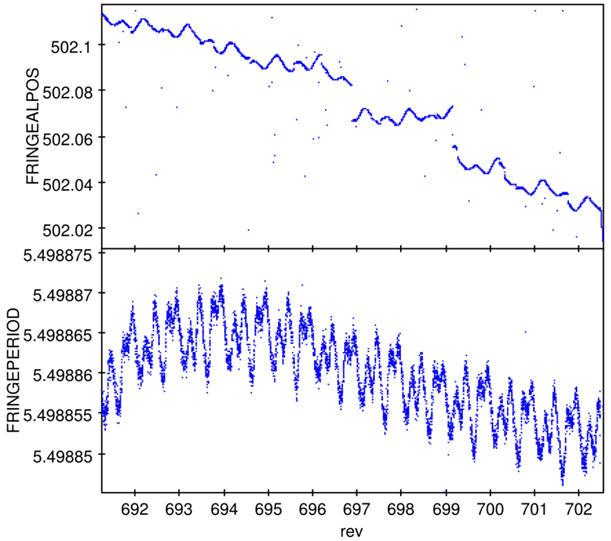}
\end{center}
\caption{BAM measurements during commissioning. Left: telescope 1, Right: telescope 2. Top: fringe phase. Bottom: fringe period. Reproduced from \cite{2014SPIE.9143E..0XM}. 
\label{fig:bamCommissioning}}
\end{figure}

Typical commissioning BAM fringe phase and period measurements were presented by \cite{2014SPIE.9143E..0XM}, and are reproduced in Fig.~\ref{fig:bamCommissioning}. Four types of variation were present at that time

\begin{enumerate}
  \item Fringe phase Sun-synchronous periodic variations with $\sim$mas amplitude. See above for their potential impact in astrometry.
  \item Fringe phase discontinuities up to mas in size and several of them per day.
  \item Fringe period variability, pseudo-periodic and with different magnitude per field of view.
  \item Fringe phase mid- to long-term evolution, mas+ in amplitude, but over time ranges of days and longer.
\end{enumerate}

Of them, the long term evolution was quickly found both wrong, the basic angle evolves but at a slower pace; and is anyway irrelevant because self-calibration handles it. However, the periodic variations of mas amplitude constitute one of the most challenging scenarios. A lot of effort was put first into determining whether the variations are real and in correcting the astrometry using the BAM measurements.

Lindegren et al. (2016, A\&A in press) explains how the basic angle variations can be expanded in terms of a Fourier series of the rotation period. Most harmonic terms have been independently derived using stellar data within the astrometric solution and ad-hoc software. Table~\ref{tab:bavVbac} shows the BAM values for the different cosine ($C$) and sine ($S$) terms together with the astrometric solution. The agreement between them is remarkable, at the level of 10-50~$\mu$as, much below the accuracy limit for Gaia Data Release 1. In addition, this demonstrates that the BAM is indeed working well and providing real basic angle measurements with unprecedented precision.

\begin{table}
\caption{Fourier coefficients $C_{k,0}$, $S_{k,0}$ determined from the BAM data compared to those obtained in a special validation run of the Gaia primary astrometric solution. Taken from Lindegren et al. (2016, A\&A in press).
\label{tab:bavVbac}}
\begin{center}
\begin{tabular}{lrrrlrr}
\hline\hline
& BAM & Solution & \quad\quad &  & BAM & Solution \\
& [$\mu$as] & [$\mu$as] &&& [$\mu$as] & [$\mu$as] \\
\hline
$C_{1,0}$ & $+865.07$ & (fixed) &&
$S_{1,0}$ & $+659.83$ & $+605.66$ \\
$C_{2,0}$ & $-111.76$ & $-134.66$ &&
$S_{2,0}$ & $-85.26$ & $-77.34$ \\
$C_{3,0}$ & $-67.84$ & $-76.14$ &&
$S_{3,0}$ & $-65.91$ & $-63.34$ \\
$C_{4,0}$ & $+18.26$ & $+24.98$ &&
$S_{4,0}$ & $+17.79$ & $+19.40$ \\
$C_{5,0}$ & $+3.20$ & $+7.42$ &&
$S_{5,0}$ & $-0.20$ & $-6.44$ \\
$C_{6,0}$ & $+3.51$ & $+6.31$ && 
$S_{6,0}$ & $+0.68$ & $+1.02$ \\
$C_{7,0}$ & $+0.03$ & $+1.45$ &&
$S_{7,0}$ & $+0.34$ & $-0.31$ \\
$C_{8,0}$ & $-0.62$ & $-2.87$ &&
$S_{8,0}$ & $-0.59$ & $-6.56$ \\
\hline
\end{tabular}
\end{center}
\end{table}

\begin{figure}
\begin{center}
\includegraphics[width=0.49\hsize]{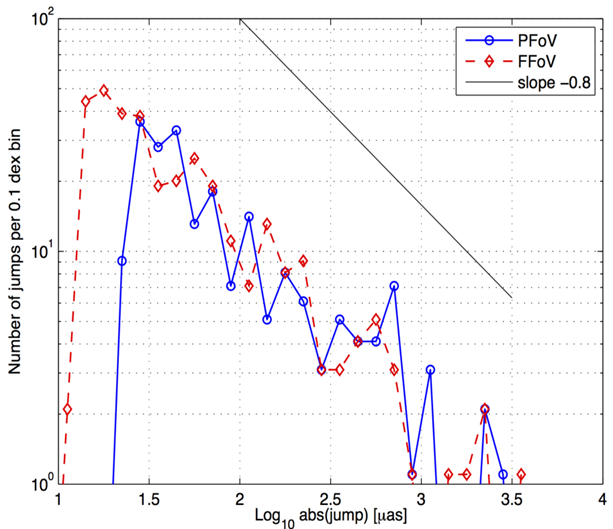}
\includegraphics[width=0.49\hsize]{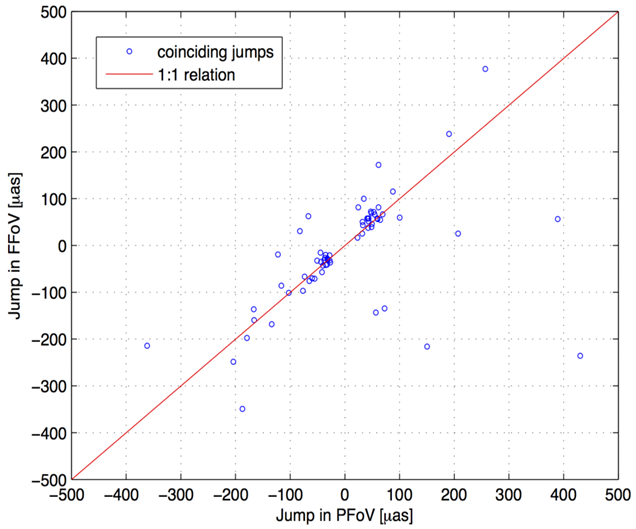}
\end{center}
\caption{BAM fringe phase discontinuities. Left: distribution as a function of the amplitude. Right: correlation between the discontinuity amplitude in both telescopes. Telescope 1: PFoV, Telescope 2: FFoV. 
\label{fig:bamDiscontinuities}}
\end{figure}

Some large BAM fringe phase discontinuities have been found real, when comparing the stellar residuals before and after the event. Routines for discontinuity automatic detection and correction have been put in place and phase jumps have been systematically identified. The average rate is around one per day during the nominal mission, and mostly concentrated after big perturbations, such as payload decontamination, refocus or safe modes. They seem to follow a power lay on the amplitude with an exponent of $\sim -0,8$ (the smaller the jump, the most probable it is). Discontinuities typically happen at the same time in both telescopes with a similar amplitude, which means that in many cases, the disturbance alters the payload geometry, but not the basic angle, at least at first order. See Fig.~\ref{fig:bamDiscontinuities} for further details.

The current sensitivity limit is around 10 $\mu$as, which means that more discontinuities can be currently hidden in the noise. Better sensitivity is expected for future BAM data reduction algorithms. Once detected, they are included in the astrometric solution. In any case, the rate of discontinuities for the payload is orders of magnitude milder than for the service module, which is consistent with the payload being much better thermo-mechanically isolated.

\begin{figure}
\begin{center}
\includegraphics[width=0.85\hsize]{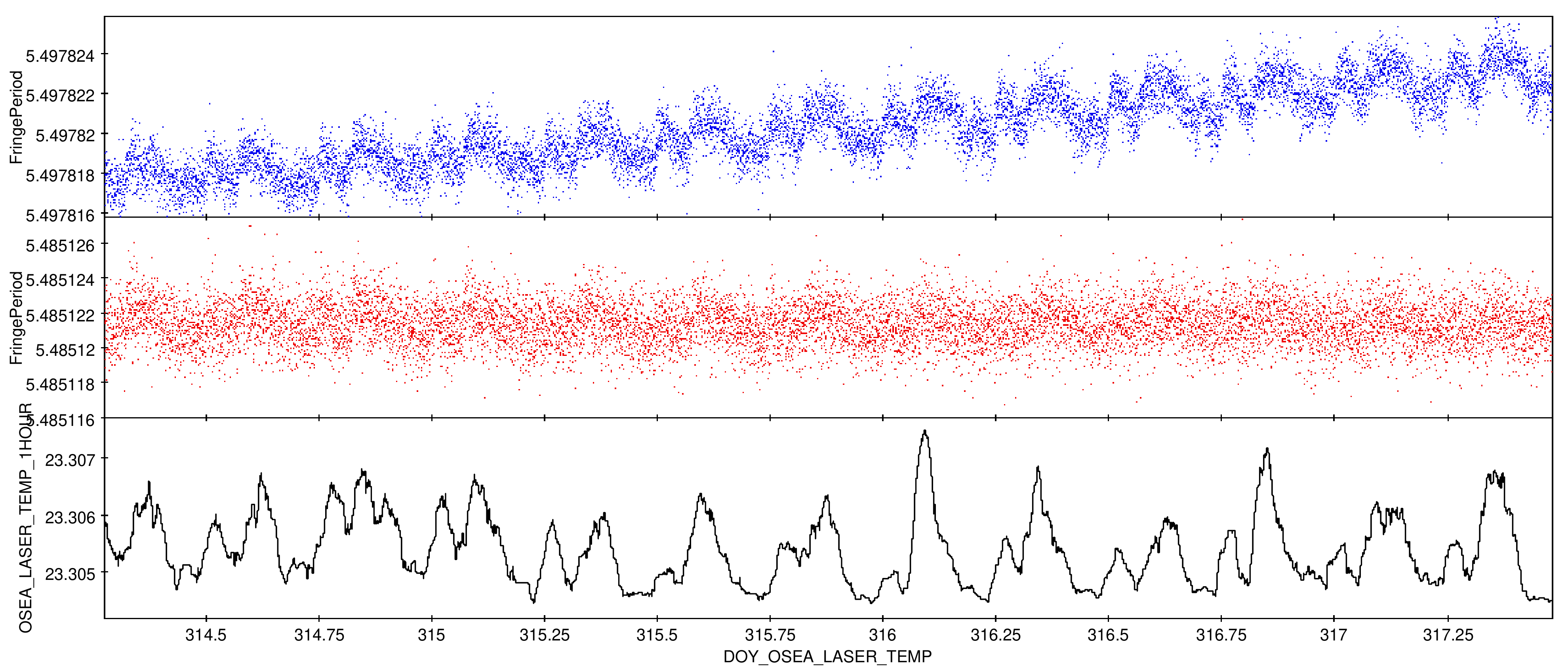}
\end{center}
\caption{BAM laser temperature (bottom) and fringe period (middle/top telescopes 1/2) evolution in a typical interval during the nominal mission. The variations are clearly correlated, although of different magnitude for each field of view.
\label{fig:bamLaserTempFringePeriod}}
\end{figure}

Fringe period variations can be relevant if the white light fringe is not exactly located in the centre of the interference pattern. In this way, fringe period variations can mimic false fringe phase variations for simple algorithms. During early commissioning, pseudo-periodic variations were identified and correlated to changes in the laser temperature \cite{2014SPIE.9143E..0XM}. The typical period stability was around $\sim$4~ppm for laser temperature changes at the level of $\pm$5~mK. However, significant improvements were found after the radial velocity spectrometer CCDs were set to only operate in high resolution mode, as a straylight mitigation countermeasure. The laser was found stable at the mK level, and the fringe period variations reduced down to $\sim$1~ppm. See Fig.~\ref{fig:bamLaserTempFringePeriod}. The changes are still different for each field of view, which is difficult to interpret as variations of the common laser wavelength. One alternative could be thermally induced focal length variations, but evidence is still unclear.

\subsection{Basic angle variations and straylight working group
\label{sect:basicAngleWorkingGroup}}

A joint ESA-Airbus DS working group was created after in orbit commissioning review to further study the origin of the straylight and basic angle variations. Very soon after its creation, the conclusions of the ESA-ESTEC members on the solar straylight, and of Airbus DS on the galactic component put an ending to the straylight investigations. The efforts were then concentrated on the study of the basic angle variations. During more than a year, many tasks were undertaken, including the analysis of big amounts of data without a-priori filtering of explanatory hypotheses, integrating the results from the Gaia astrometric data analysis, driving the refinement of the finite element thermo-elasto-optical model and finally carrying out several on-board tests around a decontamination campaign in June 2015. The final report, issued on February 2016, is a comprehensive report of all activities and results. In this subsection, only a few hints on selected aspects is presented.

\begin{figure}
\begin{center}
\includegraphics[width=0.75\hsize]{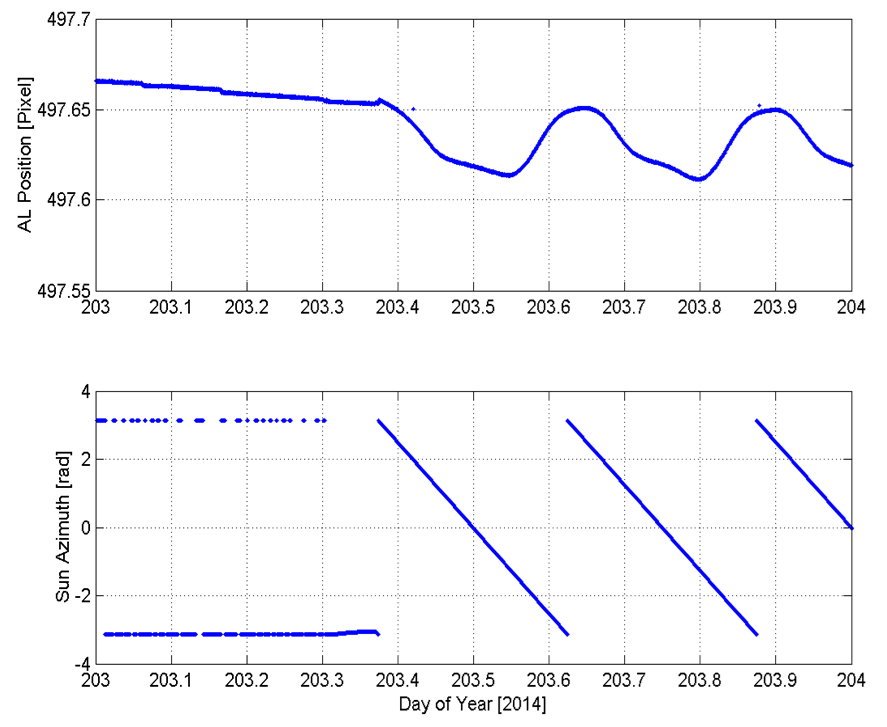}
\end{center}
\caption{Spin restart in July 2014. The basic angle variations started very soon after spacecraft spin resumed, giving support to a service module driven thermoelastic hypothesis. 
\label{fig:spinRestart}}
\end{figure}

The origin of the basic angle variations could not be traced to a single root cause. However, significant evidence suggests their origin are thermoelastic perturbations originated in the service module. As the spacecraft rotates, the temperatures in the sun-illuminated side evolve by a few degrees, inducing thermoelastic deformations. They are postulated to propagate to the payload via a yet unknown mechanism (payload and service module were effectively decoupled by design). One experimental evidence supporting this behaviour is the spin restart data, obtained in July 2014 after a safe mode was experienced. As a result, Gaia stopped spinning for some days, which halted the basic angle variations and stabilised the temperature. When scientific operations resumed, the basic angle variations restarted very soon after the rotation was initiated (within a few minutes). See Fig.~\ref{fig:spinRestart}. Only the service module could react so quickly to changes in the incoming radiation, the payload module needing more than a day to stabilise. Purely thermal or payload based hypothesis are hard to confront with this observational result.

\begin{figure}
\begin{center}
\includegraphics[width=\hsize]{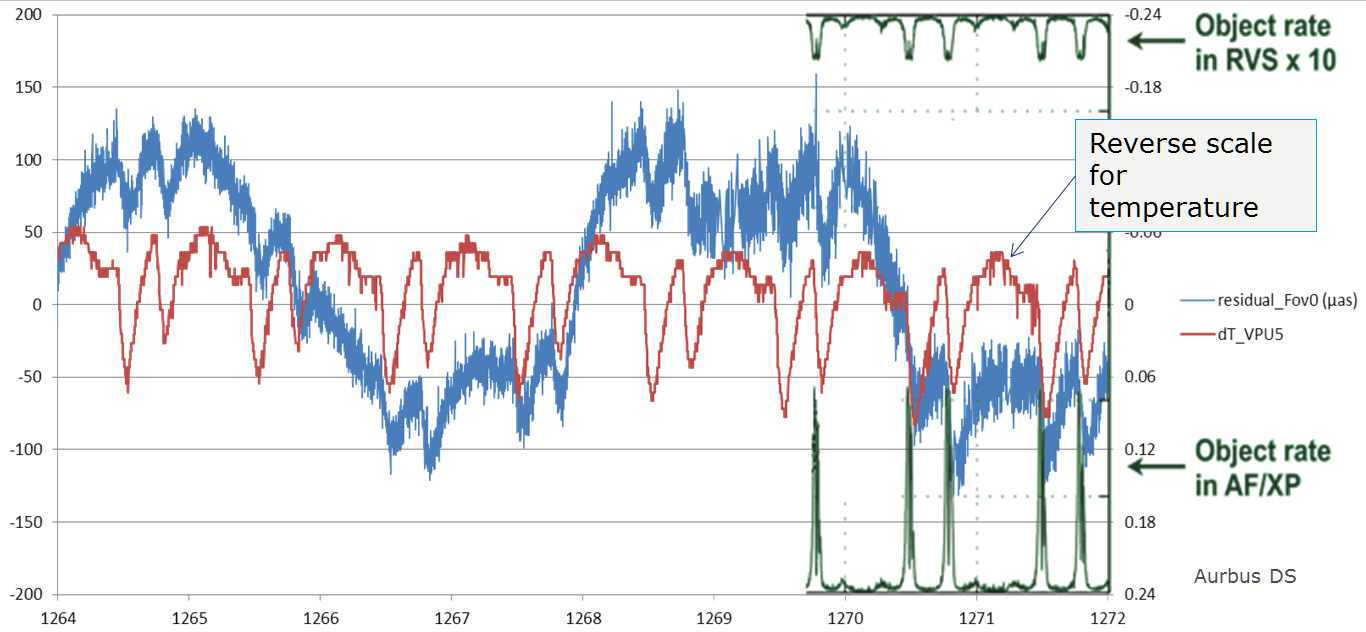}
\end{center}
\caption{Telescope 1 BAM Fourier fit residuals against the temperature of VPU5 (one computer in the service module, reversed scale) and the number of stars observed in the astrometric, photometric and spectroscopic focal planes.
\label{fig:bamFitResiduals}}
\end{figure}

When Lindegren et al. (2016 A\&A, accepted) determined the Fourier expansion of the basic angle variations evolve slowly with time, the next step was to remove that stable periodic component from the signal and inspect the residuals. Fig.~\ref{fig:bamFitResiduals} shows them for telescope 1, together with the temperature readings in one of the on-board computers (VPU5) and the number of stars observed in the astrometric and spectroscopic focal planes. Two effects are clearly visible. First, there is a $\sim$24 hours slow evolution component in addition to the six hours main period. Second, there are peaks related to the number of stars observed.

The 24 hours period was later identified as an effect of the way the downlink is operated. Even though the phased array antenna is never switched-off, the signal coding scheme changed between ground station contacts (complex signal encoding only when downlink was active). This meant the transponders consumed more power during the contacts, which are typically scheduled according to a 24 hour logic. It was decided to force signal encoding without data transmission in the antenna outside contacts (except when spacecraft ranging is needed). This action reduced the impact of the 24 hours basic variation by more than half.

The correlation with the number of stars was puzzling at first glance, due to the negligible brightness of stellar sources. However, many service module components are affected by a bigger data rate, most notably the computers and the on-board data storage. A clear correlation thus exists between e.g. VPU5 and the peak basic angle variations, giving further support to the thermoelastic hypothesis. The rule of thumb is a contribution of 0-100~$\mu$as basic angle variation per degree of thermal change amplitude for each equipment in the service module.

\begin{figure}
\begin{center}
\includegraphics[width=0.85\hsize]{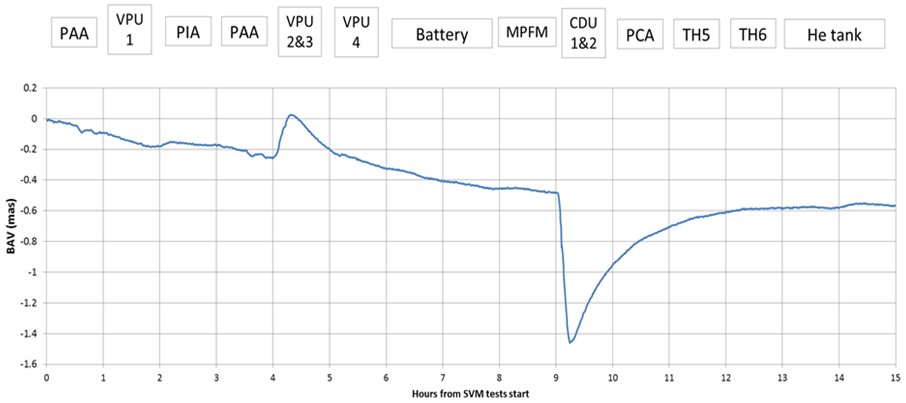}
\includegraphics[width=0.85\hsize]{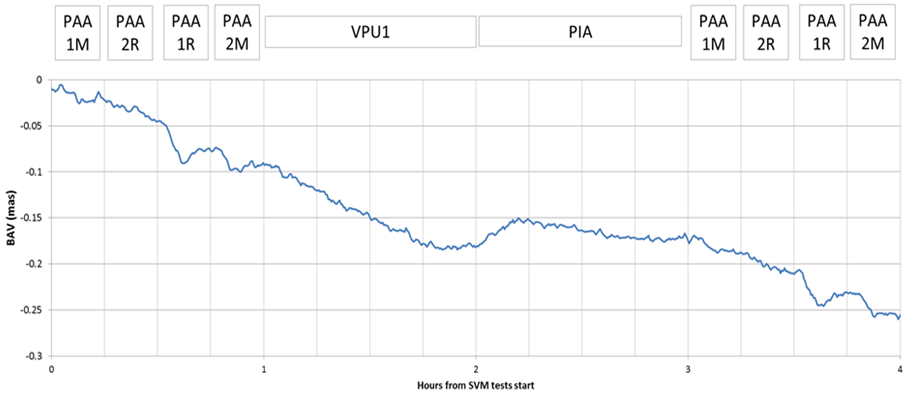}
\includegraphics[width=0.85\hsize]{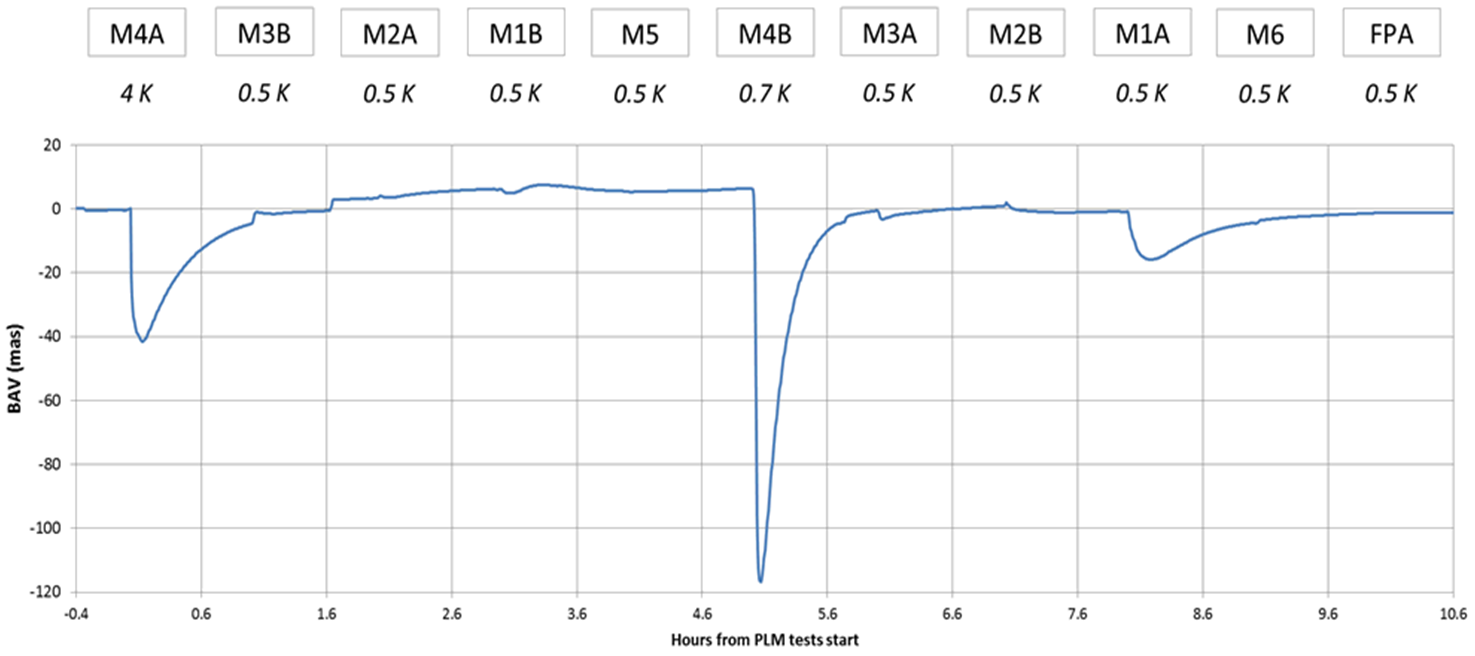}
\end{center}
\caption{Basic angle variation sensitivity to heat pulses in various service (Top and Middle) and payload (Bottom) module components. This experiment was carried out during two days before a payload decontamination event in June 2015.
\label{fig:bamSensitivities}}
\end{figure}

The normal succession of events on-board (e.g. safe modes, antenna and on-board storage switch-off, star tracker changes, ...) provided lots of data (events) to infer the influence of many systems in the basic angle variations. However, it was deemed necessary to carry out a systematic sensitivity analysis. In this way, two days of nominal mission time were devoted to ad-hoc tests just before a payload decontamination took place in June 2015. Short heat pulses were locally introduced using survival heaters throughout the spacecraft, beginning with the service module and ending with the mirrors in the payload, the temperature changes were recorded through house-keeping and confronted to the basic angle variations measured by the BAM. See Fig~\ref{fig:bamSensitivities}. This very clean data set confirmed many suspicions (impact of computers and antenna), and revealed interesting unknown effects, such as the link to the atomic clock and one mirror (M4B) in the service and payload modules, respectively. The newly found sensitivities, even if large, play a minor role, due to the very high thermal stability of these components (the effect is the product of the senstivity times the variation).

\begin{figure}
\begin{center}
\includegraphics[width=0.49\hsize]{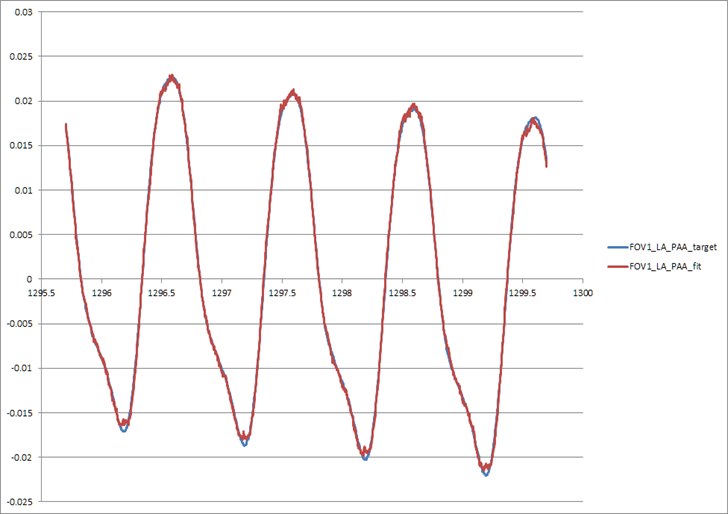}
\includegraphics[width=0.49\hsize]{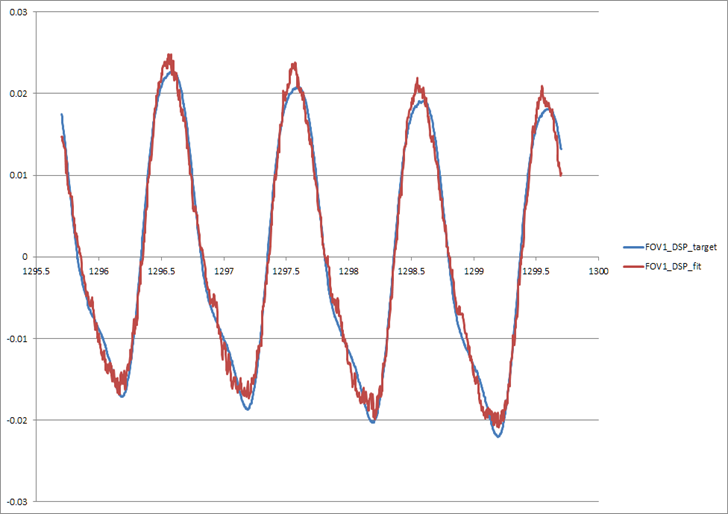}
\end{center}
\caption{BAM signal decomposition for telescope 1. Linear combination of different sets of service module temperatures can reproduce the overall shape of the basic angle variations. Left: phase-array antenna and launch vehicle adapter ring. Right: deployable sun-shield panels. 
\label{fig:bamLinearDecomposition}}
\end{figure}

Three additional and crude simplfying assumptions were added to the basic thermoelastic hypothesis to gain further insight on the origin of the perturbations. First, the variations are the result of Sun-side SVM temperature variations alone. Second, the house keeping temperatures provide a sufficient temporal and spatial sampling of the major changes (a sort of extreme macro-node analysis). Third, the BAM signal can be decomposed as the linear sum of several such temperatures via principal component analysis. No time delays were considered. Several temperature  combinations were tested. Two of them are shown in Fig.~\ref{fig:bamLinearDecomposition}. The results confirm the major features in the basic angle variations can be reproduced in terms of the  thermoelastic hypothesis. However, the decomposition is not unique, which prevents the identification of a single origin for the variations. This means such a minimalistic analysis is too simple, and the Fourier content of the service module temperature variations is the same for many systems.

\begin{figure}
\begin{center}
\includegraphics[height=4.4cm]{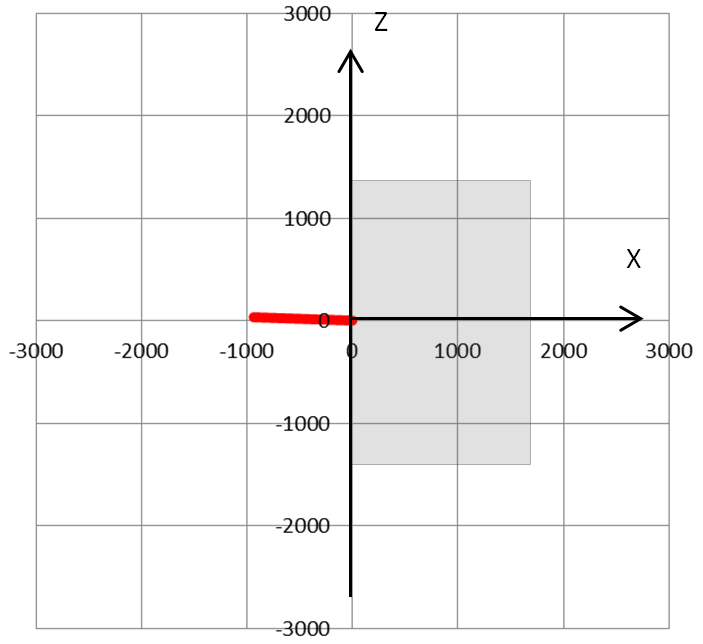}
\includegraphics[height=4.4cm]{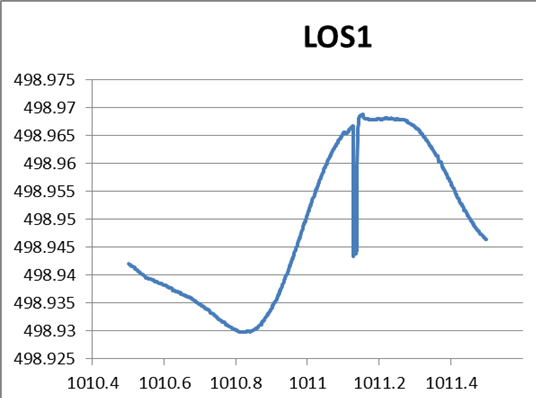}
\includegraphics[height=4.4cm]{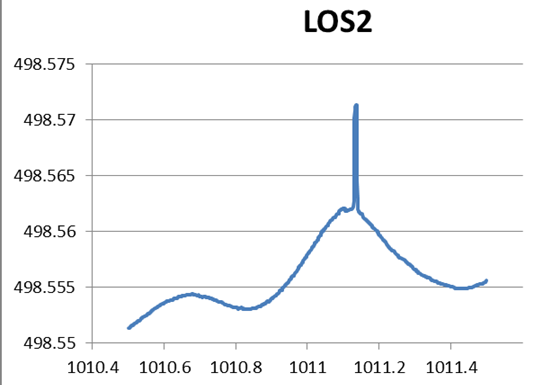}
\end{center}
\caption{Basic angle and station keeping manoeuvres. Left: schematic geometry of the forces applied to the spacecraft during one station keeping manoeuvre. Centre/Right, associated transient response in the basic angle as seen by the BAM. Plots: Airbus DS.
\label{fig:stationKeepingManouevres}}
\end{figure}

An alternative explanation for the larger than expected basic angle variations relied on the Gaia opto-elastic rigidity being much smaller than the model predictions. That is, if there is something loose within the payload. This hypothesis was tested analysing the BAM readings obtained during spacecraft station keeping manoeuvres, which are carried out routinely to maintain the orbit around the metastable L2 point. They use the chemical propulsion thrusters, which apply forces in the range $\sim$1-10~N. Fig.~\ref{fig:stationKeepingManouevres} shows an example of actuation, together with the associated transient impact on the basic angle. A careful analysis of several actuations reveals the results are basically compatible with the opto-mechanical model, and refutes a simple mechanical origin (e.g. based on inertial forces) for the basic angle variations.

\subsection{Interferogram shape and white light fringe}
\label{sect:bamShapeWhiteFringe}

The on-ground thermal vacuum tests and commissioning data showed that BAM data are more complex than the pure interference of two perfect Gaussian beams \cite{2014SPIE.9143E..0XM}. This is probably the combination of several effects, such as accumulated aberrations along the BAM optical path, partial clipping by some mirrors or interference effects within the detector itself (fringing). At the fringe level, these effects manifest as being neither purely plane-parallel nor equispaced.

\begin{figure}
\begin{center}
\includegraphics[height=6cm]{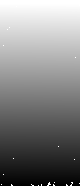}
\hspace{0.5cm}
\includegraphics[height=6cm]{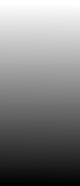}
\hspace{0.5cm}
\includegraphics[height=6cm]{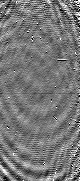}
\end{center}
\caption{BAM interferogram detailed analysis. Left: for a given telescope 1 pattern, the AL location for each fringe and AC column has been determined via cosine profile fitting, producing this staircase graded pattern. Middle: a low order 2D polynomial fit to the data seems similar at first glance. Right: row by row subtraction of the left 2D map provides the distribution of the fringe period over the interferogram, revealing significant structure.
\label{fig:bamFringeFit}}
\end{figure}

An ad-hoc processing has been carried out for selected time intervals. Basically, for each fringe and AC column, cosine fits have been carried out to determine the AL position, which is a 2D map of the fringe pattern. Row by row subtraction provides the fringe periods spatial distribution. Low order 2D polynomial fits have also been computed for comparison. Fig.~\ref{fig:bamFringeFit} shows some sample fit results for telescope 1 (the one with more elements in the BAM optical path). The inhomogeneity of the fringe period over the interferogram is evident, which raises concern on summarising it as a single and homogeneous quantity. This effect should be considered in future missions relying in very precise laser interferometry.

\begin{figure}
\begin{center}
\includegraphics[height=4.3cm]{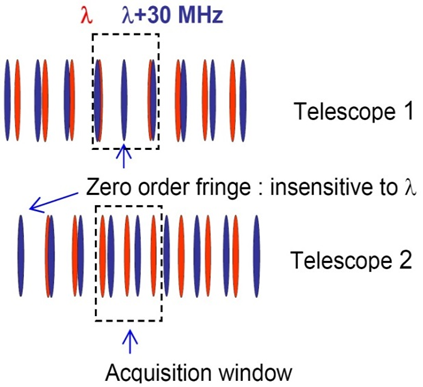}
\includegraphics[height=4.3cm]{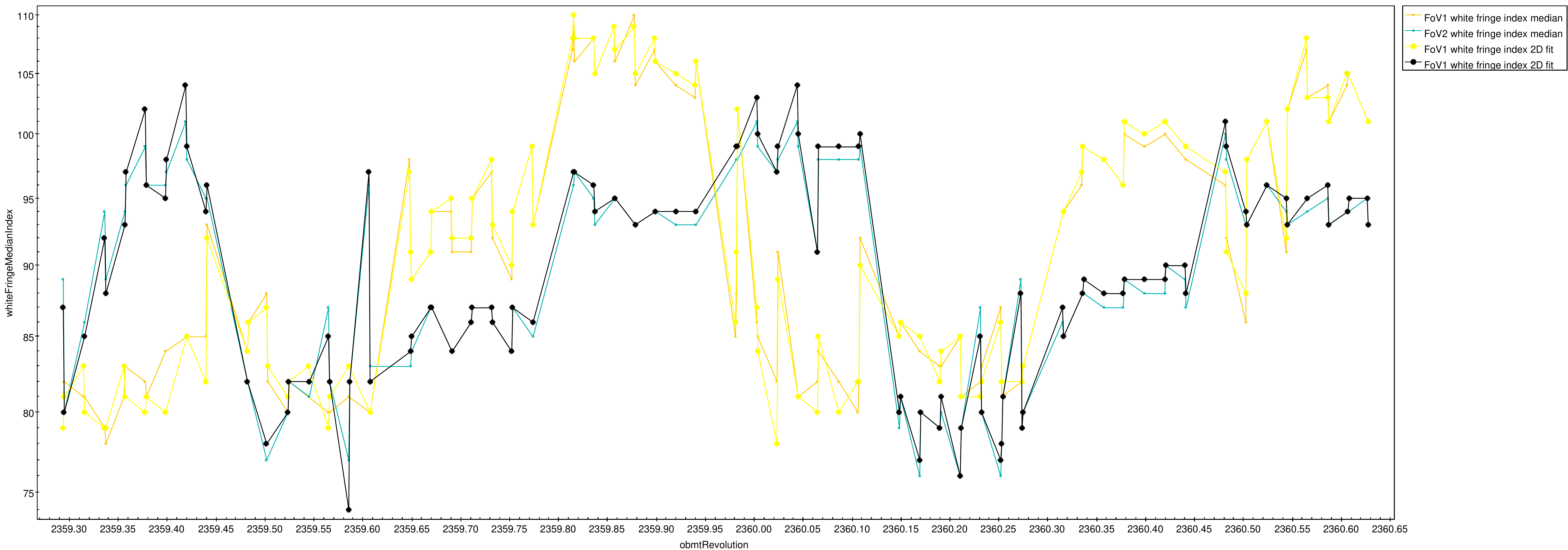}
\end{center}
\caption{BAM white light fringe. Left: effect of a decentred white light fringe plus a change in the fringe period mimicking a true pattern shift. Image: Airbus DS. Right: apparent location of the white light fringe for each field of view and several tests. 
\label{fig:whiteFringe}}
\end{figure}

There have been efforts to determine the location of the white light fringe during the on-ground thermal vacuum tests and in-orbit. Basically, the white fringe location was determined during payload integration using a non-monochromatic light source. It was thus verified that, for each interferogram, the white fringe is well placed in the central region (good control of the optical path). The reasoning behind this requirement was to avoid aliasing between true pattern shifts due to line of sight variations and the combined effect of a change in the fringe spacing and a decentered white fringe, which would provide a centre of mass shift of the interferogram, see Fig.~\ref{fig:whiteFringe}.

There have been efforts during the on-ground thermal vacuum campaigns, in-orbit commissioning and even some tests running during the nominal mission to re-determine the location of the white fringes. The objective is to verify that they are still well centred and to estimate the potential effect of aliasing with typical nominal mission fringe period changes during a revolution, which are small, $\sim$1~ppm but not zero.

The BAM has two laser sources (nominal and redundant), but no white light source. Therefore, different tests were carried out making use of the known laser wavelength sensitivity to temperature (major effect) and current intensity (minor). During the tests, a combination of two basic actions was defined and carried out: switching on and off the laser thermo-electric cooling unit and changing the current intensity over all possible pre-defined values. Noticeable changes were found in the interferogram shape. However, they were not the simple homogeneous addition of a fixed quantity to the fringe period, even if its intrinsic 2D nature was considered.

In this way, the individual fringe analysis described above was applied to the white fringe tests. For each perturbation (laser temperature or current intensity), the fringe experiencing the smaller shift in average was identified as the white light fringe. Fig.~\ref{fig:whiteFringe} shows an example of the results obtained for one particular test. Two things are evident. First, the white light fringes are indeed well centred (between fringe 75 and 110, for a total of around 180 fringes) within the interferograms. Second, there is no single fringe static against all perturbations, pointing to the changes in the laser source also having a thermal impact on the payload. Impact in the Gaia mission, if any, would be very small ($\mu$as level) and is still under investigation. The addition of white light sources should also be considered for future missions requiring Young-like interferometers.

\section{Conclusions}
\label{sect:conclusions}

Three issues related to the Gaia in-orbit stability have been reviewed in this work: the focus evolution through the mission, the additional straylight and the basic angle variations.

Regarding the telescope focus, it has been verified that Gaia provides almost diffraction limited along scan performance over a very large field of view. However, the image sharpness is not static, but evolves with time with a typical time scale of several months. The typical mitigation strategy is payload decontamination (normally driven by throughput loss), followed by a small adjustment of the telescopes, if needed. Optimum quality has always been regained afterwards. This sequence of events: increasingly long quiet interval, decontamination and refocus is expected to repeat until the end of the mission.

The sources of the excess straylight have been unambiguously identified as a stellar and solar component. Stellar parasitic light, mainly coming from the Galaxy, integrated over previously identified straylight paths is responsible for the stellar contribution. The solar straylight is the unfortunate combination of three effects: sticking out fibres in some sunshield edges, the sunshield not blocking all diffracted light onto the windows in the thermal tent, and rogue paths transporting it to the focal plane. Software mitigation measures have adopted, both on-board and on-ground.

The basic angle periodic variations and phase discontinuities identified during commissioning have been verified after a careful comparison to the stellar astrometry. The mitigation measures are the BAM measurements themselves and self-calibration within the astrometric reduction.
An ESA-Airbus DS working group was created to study the origin of the basic angle variations and straylight, discovered during commissioning by the Gaia scientists (payload experts). It revealed the detailed origins of the straylight and collected evidence suggesting the basic angle variations are thermoelastically driven from perturbations coming from the service module. The importance of having no moving parts and keeping a power load consumption profile as stable as possible has been revealed.

Some general advice for future missions requiring extreme stability has also been collected:

\begin{itemize}
  \item Design for stability, but plan for instability
  \item The state of the art in stability is mas, nm and mK 
  \item The future is $\mu$as, pm and $\mu$K. Modelling at this level (e.g. FEM) is very complex and might need additional research and new tools
  \item Some requirements are very difficult (and expensive) to verify before launch. Saving money here is typically a wrong decision
  \item Keep the spacecraft as stable and boring as possible. Routine is important
  \item House keeping is key. Appropriate precision, resolution and spatial and temporal frequency needs to be ensured, which means specialised equipment and additional downlink bandwidth
  \item High precision metrology is essential and (never too) expensive
\end{itemize}

Finally, note that Gaia is providing (and will keep doing it during the whole mission) unprecedented quality astronomical data, with the Data Release 1 scheduled for 14 September 2016.

\section{Acknowledgements}

The authors wish to thank Airbus DS for their support and access to internal documents on the wavefront sensor and Gaia optical design. Some concepts and ideas presented here come from those sources.

Material used in this work has been provided by the Coordination Units 3 (CU3) of the Gaia Data Processing and Analysis Consortium (DPAC) and ESA-Airbus DS BAM and straylight working group. They are gratefully acknowledged for their contribution.

\bibliography{2016_06_spie_focusStraylightBam,gaia_livelink_valid,gaia_livelink_obsolete,gaia_drafts,gaia_refs,gaia_books,gaia_refs_ads}   
\bibliographystyle{spiebib}   

\end{document}